\documentclass[10pt,superscriptaddress,aps,twocolumn,pra,showpacs]{revtex4-1}
\usepackage[latin1]{inputenc}
\usepackage{amsmath}
\usepackage{mathptmx}
\usepackage{graphicx}

\begin{document}

\title{Dual pumped microresonator frequency combs}

\author{T. Hansson}
\email[E-mail address: ]{tobias.hansson@ing.unibs.it}
\author{S. Wabnitz}
\affiliation{Dipartimento di Ingegneria dell'Informazione, Universit\`a di Brescia, via Branze 38, 25123 Brescia, Italy}

\pacs{42.60.Da, 42.65.Hw, 42.65.Ky, 42.65.Sf}

\begin{abstract}
A study is made of the nonlinear dynamics of dual pumped microresonator Kerr frequency combs described by a driven and damped nonlinear Schr\"odinger equation, with an additional degree of freedom in the form of the modulation frequency. A truncated four wave model is derived for the pump modes and the dominant sideband pair which is found to be able to describe much of the essential dynamical behaviour of the full equation. The stability of stationary states within the four wave model is investigated and numerical simulations are made to demonstrate that a large range of solutions, including cavity solitons, are possible beyond previously considered low intensity patterns.
\end{abstract}

\maketitle

\section{Introduction}
\label{intro}

Optical frequency comb generation using microresonators is currently a very active topic of research. Kerr frequency combs based on microresonator devices offer an intriguing alternative to mode-locked femtosecond lasers that have revolutionized the precision of optical clocks and frequency metrology \cite{kippenberg_2011}. Frequency combs can additionally be used for spectroscopy applications, and planar microresonators have the potential for enabling a new class of cheap and compact light sources that could be integrated on a chip, while at the same time providing improved sensitivity, bandwidth and acquisition time over conventional methods \cite{schliesser_2012}. However, microresonators are fundamentally nonlinear devices that are known to display a rather complicated dynamical behaviour that is not easily predictable, with both stable stationary states, corresponding to temporal patterns and cavity solitons, as well as chaotic attractors \cite{haelterman_1992c}.

Most research to date has focused on the experimental setup where a single continuous wave (CW) laser is used to pump a resonant cavity. In this paper we will instead investigate an alternative pump configuration where the cavity is simultaneously pumped at two different frequencies. Such a pumping configuration has certain benefits, allowing e.g.~the generation of frequency combs without a pump intensity threshold in both the normal and the anomalous dispersion regime. The dual pumped setup could e.g.~be particularly interesting for generating frequency combs in the normal dispersion regime and for providing stabilization of cavity solitons in resonator based optical soliton memories \cite{wabnitz_1993}. The dual pumping can be implemented either by using two separate CW laser sources, where high comb stability can be achieved by phase locking the lasers beat note to a reference oscillator \cite{strekalov_2009}, or by modulating a single CW pump laser so as to simultaneously excite two modes separated by some multiple of the free-spectral-range. The use of two pumps thus provides an additional degree of freedom in the form of the modulation frequency that is experimentally available for tuning.

Previous work on bichromatically pumped resonator cavities has primarily focused on the threshold less comb generation process and the concomitant low intensity patterns. However, it is the purpose of the article to show that the dual pumped configuration can support a rich variety of different comb states that can take the form of both stable and unstable states as the pump intensity increases. Indeed, the same physical four-wave-mixing (FWM) process that occurs for the single pump configuration is still present for dual pumping. The single pumped configuration can in fact be seen as a degenerate limit of the more general dual pumped case. It should in particular be noted that both the single pumped and the dual pumped configuration are capable of supporting cavity soliton solutions.

The dynamical evolution of the intracavity field of microresonators and whispering-gallery-mode resonators can conveniently be described using the formalism of the driven and damped nonlinear Schr\"odinger (NLS) equation \cite{matsko_2011,coen_2013}, also referred to as the Lugiato-Lefever equation. This model has previously been used to describe the dynamical evolution of dispersive fiber-ring cavities, where it arises as the mean-field limit of the discrete Ikeda map \cite{haelterman_1992}. The dynamical evolution can alternatively be modeled by using an approach based on frequency-domain coupled mode equations for each resonant mode \cite{matsko_2005,chembo_2010}. Both of these formalisms have recently been shown to provide equivalent descriptions for the evolution of the optical field and provide complimentary views on the comb generation process \cite{chembo_2013}. The coupled mode equations provide a spectral approach that models the slow temporal evolution of the optical field using a system of coupled ordinary differential equations for each mode, while the driven and damped NLS equation is a time-domain approach that models the field using a single partial differential equation with two separate time-scales for the evolution and for the temporal profile. Both of the formalisms can be efficiently simulated using Fast Fourier Transform methods such as those used in the split-step Fourier method. This numerical method propagates the field in the frequency-domain, while the nonlinear contribution is calculated in the time-domain, and permits the evolution of wideband frequency combs to be numerically simulated with equal efficiency using either formalism \cite{NCME}.

The two theoretical formalisms for the comb generation process are both infinite dimensional and rather involved. Valuable insight into the comb generation process can readily be obtained by instead considering finite mode truncations of these systems \cite{TWM}. The motivation for this is that a significant amount of the total energy of the system is contained in just a few modes that can be modeled using a low dimensional system of ordinary differential equations. In this paper we consider a truncated four wave model consisting of two identical strong pump modes and the dominant sideband pair. Truncated models are useful because they can often capture the essential dynamical behaviour such as dynamically attracting states of the full system. Finite mode models frequently show remarkably good agreement with numerical simulations, which demonstrates their utility in addition to providing physical understanding. The truncated model can in particular be used to find parameter regimes for stationary comb states, and also to investigate their range of stability.

In the next section \ref{model} we will derive the dynamical four wave model and the equations for the stationary states that correspond to its fixed points. In section \ref{dualcombs} we then compare the stationary states to numerical simulations of the full system and give characteristic examples of the range of dynamical behaviour that can be observed for dual pumped frequency combs. Finally in section \ref{conclusions}, we discuss the agreement of the model with the numerical simulations and summarize the results.

\section{Four wave model}
\label{model}

The formalism we will use in order to model a bichromatically pumped microresonator is the mean-field formalisms of the driven and damped NLS equation. To simplify the analysis we consider a restricted case where the two pumps have equal amplitudes and phases as well as a common phase detuning, in which case the intracavity field can, to the lowest order of dispersion, be described by the following normalized equation:
\begin{equation}
    \frac{\partial A}{\partial\tau} + i\beta\frac{\partial^2 A}{\partial t^2} - i|A|^2A = -\left(1+i\delta_0\right)A + f_0\cos(\Omega t).
    \label{eq:ddNLS}
\end{equation}
Where $t$ is the ordinary (fast) time variable which describes the temporal profile of the field, and $\tau$ is the slow time-scale for the evolution of this profile over successive round-trips. The field is moreover assumed to be periodic with the cavity round-trip time, so that $A(t+2\pi,\tau) = A(t,\tau)$ \cite{TWM}. The remaining parameters in the equation are defined as follows: $\beta$ is the second-order dispersion coefficient, $\delta_0$ is the cavity detuning, $f_0$ is the external pump field and $\Omega$ is the pump modulation frequency. There is thus an additional free parameter compared to the single pump case which is described by the Lugiato-Lefever equation that is obtained from the above equation in the limit of $\Omega \to 0$. We emphasize that Eq.(\ref{eq:ddNLS}) is a rather general model that can be valid also for other physical systems beyond microresonators, such as e.g. dispersive fiber-ring resonators \cite{haelterman_1992}.

Because of the detuning parameter that is not present for the ordinary NLS equation, one finds that modulation instability (MI) can occur both for normal and anomalous dispersion in microresonators, allowing phase matching of pump modes and sidebands also in the normal dispersion regime \cite{haelterman_1992b,matsko_2012}. The degenerate MI process is the primary comb generation mechanism for the single pumped case and has a gain spectrum for the driven and damped NLS equation that is somewhat different from that of the ordinary NLS equation, being usually peaked quite far from the pump mode while not providing any gain for frequencies close to the pump \cite{TWM}. Note that it is assumed in Eq.(\ref{eq:ddNLS}) that the difference in detuning between the two pumps is sufficiently small to be negligible compared to the average (common) detuning. The effect of the difference is otherwise to introduce a phase drift on the slow time-scale between the two pumps, which may have a detrimental effect on the comb generation process if it becomes too large.

In order to derive the four wave model we assume that the field can be expanded using a symmetric four wave ansatz, c.f. Ref.~\cite{trillo_1994}
\begin{equation}
    A(t,\tau) \approx A_p(t,\tau)\cos(\Omega t) + A_s(t,\tau)\cos(3\Omega t).
\end{equation}
It should be noted that this ansatz requires that $|\Omega| > 0$, so that it does not cover the degenerate case of the three wave model. Collecting terms with the frequency dependence $\exp(\pm i\Omega t)$ and $\exp(\pm i3\Omega t)$ while neglecting higher order terms results in a four dimensional system of coupled mode equations, which can be rewritten more conveniently by introducing a new set of dynamical variables, viz.
\begin{align}
    & \eta = |A_p|^2/P_0, & \phi = \phi_s - \phi_p, \nonumber\\
    & P_0 = |A_p|^2+|A_s|^2, & \theta = \phi_{f_0} - \phi_p.
\end{align}
These variables are similar to those of the three wave model \cite{TWM} and correspond to the normalized pump mode intensity $\eta$, the relative phase between pump mode and sidebands $\phi$, the total intensity $P_0$ and the relative phase between the external pump and the pump mode $\theta$. Defining a new dispersion parameter $\kappa = \beta\Omega^2$ we find this set of variables to be governed by the following system of equations that describes the nonlinear dynamics of the truncated four wave model for the pump modes and the dominant sideband pair:
\begin{widetext}
\begin{align}
    & \frac{\partial\eta}{\partial\tau} = -2\eta(1-\eta)\left(P_0 h_1(\eta,\phi)\sin\phi - \frac{|f_0|}{\sqrt{P_0\eta}}\cos\theta\right), \label{eq:eta}\\
    & \frac{\partial\phi}{\partial\tau} = \left(8\kappa-\frac{P_0}{4}\right) + P_0\eta h_1(\eta,\phi)\cos\phi - P_0(1-\eta)h_2(\eta,\phi) - \frac{|f_0|}{\sqrt{P_0\eta}}\sin\theta, \label{eq:phi}\\
    &  \frac{\partial P_0}{\partial\tau} = -2P_0 + 2|f_0|\sqrt{P_0\eta}\cos\theta, \label{eq:P0}\\
    & \frac{\partial\theta}{\partial\tau} = \left(\delta_0 - \kappa - P_0\right) + \frac{P_0\eta}{4} - P_0(1-\eta)h_2(\eta,\phi) - \frac{|f_0|}{\sqrt{P_0\eta}}\sin\theta. \label{eq:theta}
\end{align}
\end{widetext}
with $h_i$ being functions of $\eta$ and $\phi$ defined as
\begin{equation}
    h_1(\eta,\phi) = \cos\phi+\frac{1}{4}\sqrt{\frac{\eta}{1-\eta}},
\end{equation}
\begin{equation}
    h_2(\eta,\phi) = \frac{1}{2}\cos(2\phi)+\frac{3}{4}\sqrt{\frac{\eta}{1-\eta}}\cos\phi.
\end{equation}
In order to find the stationary states of this system it is helpful to eliminate the two phase variables $\theta$ and $\phi$. One then obtains an algebraic system of coupled equations for the pump mode intensity $I_0 = P_0\eta/2$ and the normalized pump mode intensity $\eta = 1/(1+x^2)$, viz.
\begin{equation}
    \frac{|f_0|^2}{2I_0} = \left[\left(\delta_0-\kappa-\frac{3I_0}{2}\right) - g_+ x^2\right]^2 + (1+x^2)^2,
    \label{eq:S1}
\end{equation}
\begin{equation}
    9\kappa = \delta_0 - g_- -\frac{3I_0x^2}{2}
    \label{eq:S2}
\end{equation}
where
\begin{equation}
    g_\pm = I_0\left[1 + \cos\phi\left(2\cos\phi+\frac{1}{2x}\left(2\pm1\right)\right)\right]
    \label{eq:gpm}
\end{equation}
and
\begin{equation}
    \cos\phi = \frac{\delta_0-9\kappa-I_0-\frac{3I_0x^2}{2}}{\sqrt{\left(\delta_0-9\kappa-I_0-\frac{3I_0x^2}{2}\right)^2+1}}.
    \label{eq:cos_phi}
\end{equation}
The two coupled equations (\ref{eq:S1}-\ref{eq:S2}) are to be solved simultaneously in order to find the fixed points of the four wave model, and the stability of the comb states is ascertained by linearizing the system (\ref{eq:eta}-\ref{eq:theta}) around the fixed points and determining whether the real parts of the eigenvalues are negative. This allows us to determine possible equilibrium states that may correspond to dynamical attractors of the original Eq.(\ref{eq:ddNLS}). It should however be noted that even if an equilibrium state is predicted to be stable by the truncated model it is not necessarily a stable state of the full equation. Stable states of the four wave model can correspond to not only stationary patterns and cavity solitons, but also to dynamical equilibrium states such as e.g.~breather states involving higher order sidebands. However, as long as the assumptions of the four wave model are satisfied one should not expect to find that stationary states appears where the truncated model predicts instability.

\section{Dual pumped frequency combs}
\label{dualcombs}

Previous work on dual pumped frequency combs has mainly focused on low intensity patterns. An experimental demonstration of bichromatically pumped frequency comb generation for a similar system to Eq.(\ref{eq:ddNLS}) was presented in Ref.~\cite{strekalov_2009} using a passive ring cavity. It was found that the comb generation process occurs without a pump intensity threshold, and the coupled mode formalism was used to derive an approximate stationary solution for the low intensity case when the sideband amplitudes are small with respect to the pump mode amplitudes.

The threshold less comb generation process can easily be understood with reference to Eq.(\ref{eq:ddNLS}). In the case that two strong pump modes are simultaneously present in the cavity it is no longer possible to find a stationary solution involving only the two pump mode frequencies. Substitution of an ansatz $A(t,\tau) = a_1\exp(i\Omega t) + a_2\exp(-i\Omega t)$ into Eq.(\ref{eq:ddNLS}) shows instead that new frequency components are generated through the nonlinear term. The pump field therefore does not correspond to an equilibrium state of the cavity as it is for the pump field of the single pumped configuration which corresponds to the stationary CW solution. The primary comb generation mechanism is consequently no longer the degenerate and intensity dependent MI process, but the non-degenerate and threshold less FWM of the two pump modes. The dual pumped configuration thus allows frequency combs generation both for normal and anomalous dispersion, even when the single pumped configuration would correspond to a stable CW solution. However more complicated states also appear as the pump intensity increases. Indeed, the same comb generation mechanisms that are of importance for the single pumped case are also important for dual pumping.

\begin{figure}[htb]
  \centering
  \includegraphics[width=0.9\linewidth]{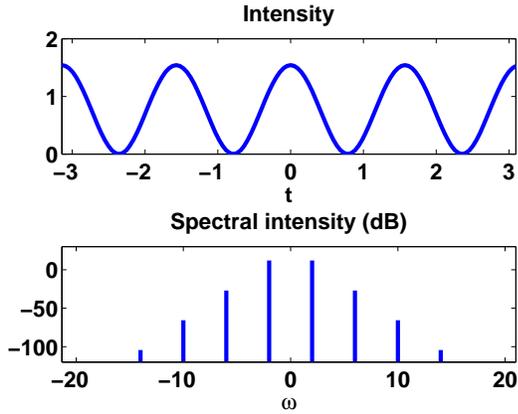}
  \caption{(Color online) Normal dispersion frequency comb for a dual pump configuration. Numerical simulation of Eq.(\ref{eq:ddNLS}) showing a stationary background free comb corresponding to a pattern of four pulses for parameters $\kappa = 1$, $\delta_0 = 3$, $|f_0| = 3$ and $\Omega = 2$.}
  \label{fig:dual2}
\end{figure}

\begin{figure}[htb]
  \centering
  \includegraphics[width=0.9\linewidth]{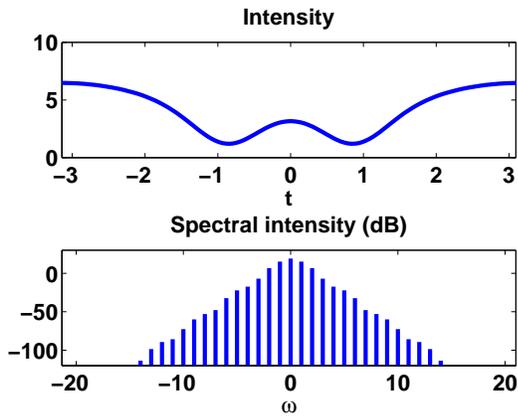}
  \caption{(Color online) Normal dispersion frequency comb for a dual pump configuration with a finite background component owing to the growth of a secondary comb which fills in the comb spectrum. Parameters are $\kappa = 1$, $\delta_0 = 5$, $|f_0| = 5$ and $\Omega = 1$.}
  \label{fig:dual1}
\end{figure}

The dual pumped frequency combs that are generated below the MI threshold are stable, and also background less since they do not have a DC component. The temporal profile of these combs is a stationary pattern of pulse shaped objects whose number can be controlled by changing the modulation frequency, with the periodicity of the temporal pattern being equal to twice the normalized modulation frequency, i.e.~the separation distance in free-spectral-ranges between the two pump frequencies. A modulation frequency $\Omega = 1$ will thus give rise to a pattern with two pulses while $\Omega = 2$ will give rise to a pattern with four etc. This is demonstrated by the two stable soft excitation frequency combs shown in Figs.~\ref{fig:dual2}-\ref{fig:dual1} for the case of normal dispersion. Note that the pump intensity is sufficiently large in Fig.~\ref{fig:dual1} to cause the growth of a secondary comb through MI which fills in the modes, including the DC component, with even mode numbers. The dual pumped combs that generate a DC background component are generally found to be unstable in the anomalous dispersion regime, while they are generally, but not always, found to be stable for normal dispersion.

\begin{figure}[htb]
  \centering
  \includegraphics[width=0.9\linewidth]{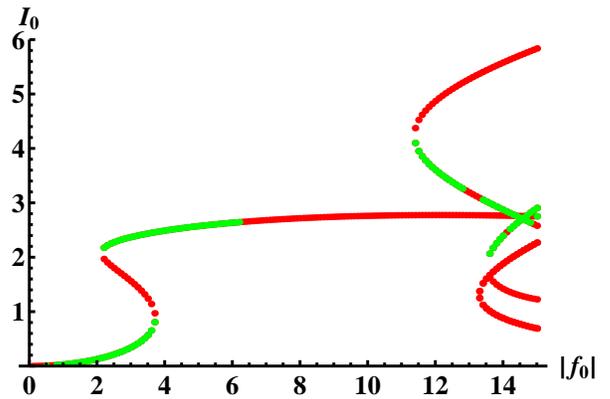}
  \caption{(Color online) Fixed point curve showing pump mode intensity as a function of external pump amplitude for fixed dispersion $\kappa = -1$ and detuning $\delta_0 = 3$. Green (light gray) curves are stable fixed points while red (dark gray) curves are unstable fixed points.}
  \label{fig:f0}
\end{figure}

\begin{figure}[htb]
  \centering
  \includegraphics[width=0.9\linewidth]{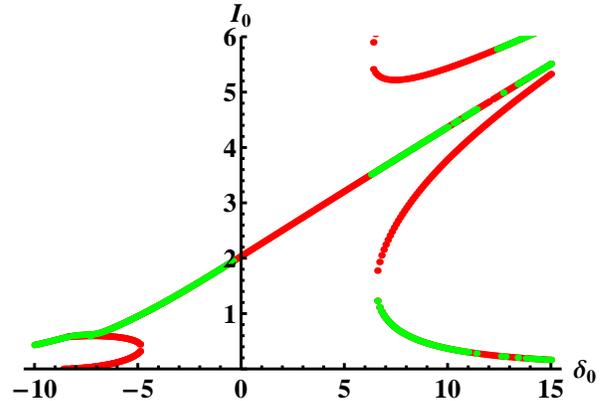}
  \caption{(Color online) Fixed point curve showing pump mode intensity as a function of detuning for fixed dispersion $\kappa = -1$ and external pump amplitude $|f_0| = 9$. Green (light gray) curves are stable fixed points while red (dark gray) curves are unstable fixed points.}
  \label{fig:delta0}
\end{figure}

The four wave model predicts the appearance of new stationary states as the driving strength of the external pump intensity increases. Some examples for the case of anomalous dispersion are shown in Figs.~\ref{fig:f0}-\ref{fig:delta0}. These show fixed point curves that are obtained by solving Eqs.(\ref{eq:S1}-\ref{eq:S2}) for a given dispersion and modulation frequency $\kappa = \beta\Omega^2$ while varying either the detuning or the external pump amplitude. The fixed point curves show that the system (\ref{eq:eta}-\ref{eq:theta}) only has a single stationary solution as the external pump intensity $f_0$ tends to zero, and that it exhibits a bistable behaviour similar to that of the CW solution which manifests itself by the characteristic S-curve dependence of the pump mode intensity when it is plotted as a function of the external pump amplitude for a fixed detuning, see Fig.~\ref{fig:f0}. This figure also shows that there exists a region of parameters (for $|f_0|$ between about 6.5 and 11) for larger pump amplitudes where no stable solutions are present. In fact the upper fixed point curve is generally found, from numerical simulations of the full Eq.(\ref{eq:ddNLS}), to be unstable also for lower values of the external pump amplitude even when the four wave model predicts stability. This is due to the influence of higher order sidebands that are neglected within the truncated model.

\begin{figure}[htb]
  \centering
  \includegraphics[width=0.9\linewidth]{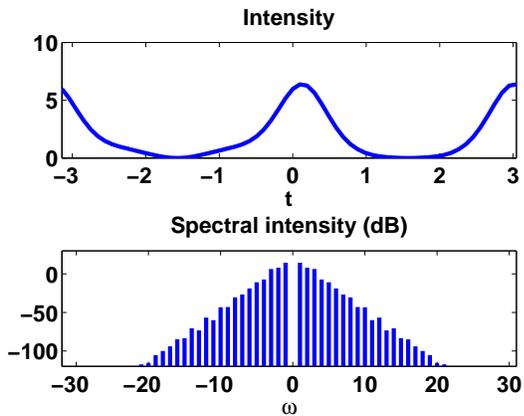}
  \caption{(Color online) Stable anomalous dispersion frequency comb without a background component but with even modes filled in due to secondary comb generation. The dispersion parameter $\beta = -1$, the modulation frequency $\Omega = 1$ and the external pumping is $|f_0| = 9$, while the detuning has a negative value $\delta_0 = -2$.}
  \label{fig:DC1}
\end{figure}

Nevertheless, we find that the truncated model predicts a number of different stationary comb states that are indeed stable. A nontrivial example of a stable anomalous dispersion comb is seen in Fig.~\ref{fig:DC1} which shows a stationary frequency comb state obtained for $\kappa = -1$ and negative detuning $\delta_0 = -2$, where a secondary comb has filled in the even modes between the odd modes of the primary comb with the exception of the DC component. The resulting temporal intensity profile are two slightly distorted pulses that are shown in the upper part of the figure. If the detuning is further increased then the secondary comb generation will still occur but the DC component will no longer be suppressed, which results in instability of the frequency comb.

It is instructive at this point to consider the various mechanisms for frequency comb generation in order to highlight the differences between single and dual pumped cavities. The primary comb generation mechanism for a single pumped cavity is the modulation instability (MI) of the pump mode. This MI process is responsible for creating the primary sidebands, which are usually located multiple free-spectral-ranges away from the pump mode frequency. The primary sidebands can in turn interact with the pump mode to create additional sidebands through the four-wave-mixing process. The sidebands can also, depending on the external pump intensity and detuning, become sufficiently strong so that they may themselves experience MI and create overlapping subcombs. Comb generation can further occur through cascaded FWM that creates new sidebands even further away from the pump mode and that can produce ultra wideband combs, which may even span a full octave as has been demonstrated by experiments.

\begin{figure}[htb]
  \centering
  \includegraphics[width=0.9\linewidth]{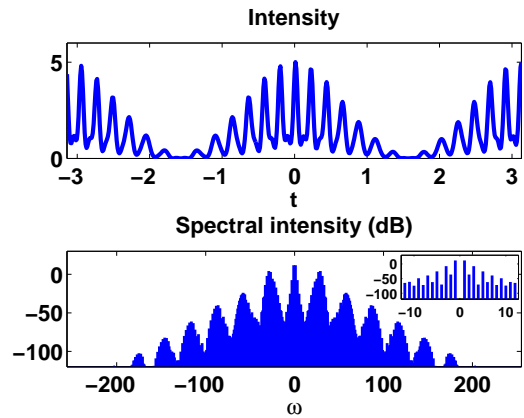}
  \caption{(Color online) Numerical simulation of Eq.(\ref{eq:ddNLS}) showing cascaded comb generation in the anomalous dispersion regime. The comb is stable and the inset shows that the DC component is missing. Parameters are $\beta = -0.01$, $\Omega = 1$, $\delta_0 = -5$ and $|f_0| = 9$.}
  \label{fig:cascaded}
\end{figure}

The cascaded comb generation is found to occur for small values of the dispersion parameter. This can be seen in Fig.~\ref{fig:cascaded} for $\kappa = -0.01$ where an anomalous dispersion frequency comb comprised of about 400 modes with an intensity larger than $-120~dB$ is shown. The cascaded comb is clearly seen to be made up of several smaller subcombs that have merged to form a broad but uneven comb without a DC component as shown by the zoomed inset. The corresponding temporal profile is a modulated train of pulses as can be seen in the upper part of the figure.

\begin{figure}[htb]
  \centering
  \includegraphics[width=0.9\linewidth]{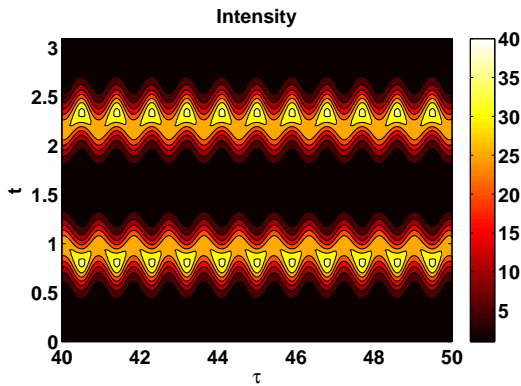}
  \caption{(Color online) Contour plot showing evolution of a stable hard excitation breather state with two oscillating bright pulses. Lighter color indicates higher intensity. Parameters are $\kappa = -1$, $\Omega = 1$, $\delta_0 = 14$ and $|f_0| = 9$.}
  \label{fig:breather}
\end{figure}

One of the main advantages of the nonlinear four wave model is that it also predicts the existence of hard excitations, i.e. comb states that cannot be reached by adiabatic changes of the external pump power and/or detuning when starting from initial conditions corresponding to an otherwise empty cavity \cite{matsko_2012b}. A soft excitation is conversely any stationary comb state that can be reached by slow changes of the pump parameters when starting from zero initial conditions. The basin of attraction for a hard excitation state is such that it requires initial conditions where multiple sidebands are already present, or such states may alternatively be reached by applying rapid non-adiabatic changes to either pump power or detuning. A hard excitation state can be interpreted as an isolated attractor in an infinite dimensional phase space of possible comb states that is not connected to the state that corresponds to zero initial conditions by any trajectory. The hard excitation states are consequently not connected to any of the low intensity patterns that can be generated through FWM of the pump modes for arbitrary parameters and that are predicted from linearized analysis. These should instead be considered as soft excitations. Nontrivial soft excitation comb states featuring e.g.~secondary comb generation can occur for both normal and anomalous dispersion, with Fig.~\ref{fig:dual1} being an example of a soft excitation for normal dispersion and Fig.~\ref{fig:DC1} an example for anomalous dispersion. Numerical simulations of Eq.(\ref{eq:ddNLS}) shows that the stable hard excitation states that can be seen in Figs.~\ref{fig:f0}-\ref{fig:delta0}, e.g.~for parameters $\kappa = -1$, $\Omega = 1$, $\delta_0 = 14$ and $|f_0| = 9$ correspond to a breather states of two bound oscillating bright pulses as seen in Fig.~\ref{fig:breather}, and similar breather states of dark pulses can also be found for normal dispersion.

\begin{figure}[htb]
  \centering
  \includegraphics[width=0.9\linewidth]{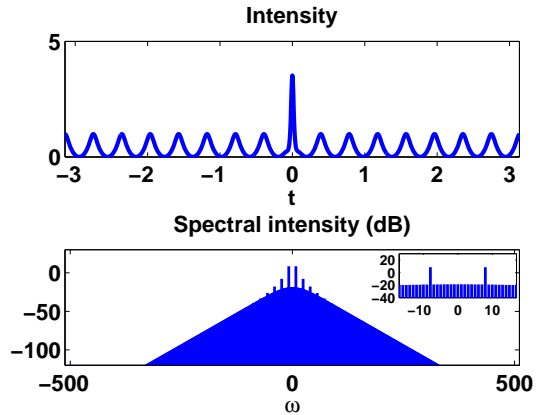}
  \caption{(Color online) Numerical simulation of Eq.(\ref{eq:ddNLS}) showing coexistence of a single injected cavity soliton on top of a stationary 16 bit pattern of low intensity clock pulses. Anomalous dispersion with $\beta = -0.001$, modulation frequency $\Omega = 8$, detuning $\delta_0 = 1.7$ and external pump amplitude $|f_0| = 1.3$.}
  \label{fig:CS}
\end{figure}

\begin{figure}[htb]
  \centering
  \includegraphics[width=0.9\linewidth]{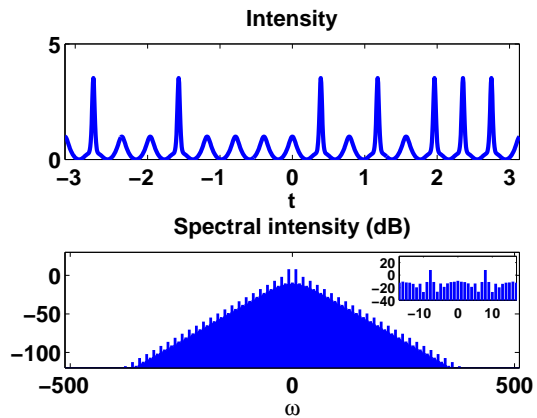}
  \caption{(Color online) Numerical simulation of Eq.(\ref{eq:ddNLS}) showing coexistence of multiple injected cavity soliton on top of a stationary 16 bit pattern of low intensity clock pulses. Same parameters as in Fig.~\ref{fig:CS}}
  \label{fig:CS2}
\end{figure}

One of the most interesting properties of the dual pumped configuration is that is also capable of supporting cavity solitons. The cavity solitons are isolated pulses that exists on a background, and that have a smooth spectral profile that involves all cavity modes with the fundamental mode spacing of a single free-spectral-range. Cavity solitons are subcritical structures that appear just below the intensity threshold for MI, where the CW background in the single pumped case or the low intensity pattern for the dual pumped case is converted into a train of pulses. This pulse train is however not composed of cavity solitons since it does not have the fundamental mode spacing but rather the mode spacing of the pump, i.e.~twice the modulation frequency. The cavity solitons can be created either by rapidly changing the pump parameters, or more easily by injecting a similar writing pulse into the cavity. The latter scheme were an addressing pulse is used to write a cavity soliton at a specific time-slot in the cavity is the basis for an optical soliton buffer which has recently been experimentally demonstrated in a fiber-ring resonator \cite{leo_2010}. Resonator based optical memories that use optical cavity solitons as bits could potentially benefit from using a bichromatic pumping scheme to suppress interactions between individual solitons \cite{wabnitz_1993}. The pumping of two modes that are separated by multiples of the free-spectral-range introduces a low intensity pattern of non-interacting clock pulses whose number can be controlled by the modulation frequency in order to determine the number of bits in the cavity. These clock pulses provide a fixed background onto which the cavity solitons can be written, and the modulation frequency can be selected so that only a single cavity soliton will fit onto each clock pulse. The clock pulses could additionally be used to provide a re-timing feature in order to synchronize detection. The same pulses that are used to write the cavity solitons can also be used to erase an optical bit for a proper choice of amplitude and width, thus allowing the cavity to function as a logical XOR gate. The entire pattern can moreover by erased by lowering the pump amplitude below the threshold for cavity solitons. An example of a single cavity soliton that has been injected into the cavity to sit on top of a 16 bit clock pattern obtain for the modulation frequency $\Omega = 8$ is shown in Fig.~\ref{fig:CS}. The corresponding frequency comb has about 600 modes with an intensity larger than $-120~dB$ which is somewhat broader than the cavity soliton comb for the single pumped case. Another example featuring a pattern of multiple cavity solitons is shown in Fig.~\ref{fig:CS2}. The spectrum is modulated and has a somewhat higher average spectral intensity compared to that of Fig.~\ref{fig:CS}, with neither spectrum showing any depression of the intensity around the pump modes, c.f.~e.g.~Fig.~2 of Ref.~\cite{chembo_2013}. It should also be noted that the single FSR comb is only present when there is a mix of high and low amplitude states. The single FSR comb is thus present as long as at least one, but not all, of the background pulses is in their excited cavity soliton state.

\section{Conclusions}
\label{conclusions}

In this article we have studied some aspects of the nonlinear dynamics of Kerr frequency comb generation in bichromatically pumped microresonator devices. The dual frequency pumping offers an alternative to the commonly used single pump setup and displays a different dynamical behaviour than the latter, allowing e.g.~the generation of frequency combs without an intensity threshold in both the normal and the anomalous dispersion regime. We have specifically considered a case that can be described within the formalism of a driven and damped nonlinear Schr\"odinger equation where the two pumps have identical amplitudes, phases and a common detuning. The two pumps then provide an additional degree of freedom that is experimentally available in the form of the modulation frequency. To gain additional insight into the dynamics we have derived a truncated four wave model for the two strong pump modes and the modes of the dominant sideband pair, which allows us to capture much of the essential behaviour of the driven and damped NLS equation within a four dimensional system of coupled equations. The analysis is fully nonlinear and allows us to predict the existence of new stationary states beyond the low intensity patterns that have previously been studied. The stationary states are found by considering fixed points of the four wave model, and their stability is investigated in order to search for stable stationary states that may correspond to dynamical attractors of the driven and damped NLS equation. However, it should be kept in mind that the truncated four wave model relies on the assumption that any additional frequency components are small, which means that stability of stationary states within the four wave model is no guarantee for the stability of the corresponding state for the full equation. By comparison with numerical simulation we find the four wave model to be very useful for predicting stability and parameters regimes of different comb states, even though the agreement is somewhat worse than for the three wave model studied in Ref.~\cite{TWM}. This is primarily due to neglecting the DC component, which may sometimes become larger than the two pump modes.

We have additionally discussed some of the mechanisms for comb generation and made numerical simulations of different frequency comb states that are possible for a dual pump configuration and that have been predicted by the four wave model. We have demonstrated that secondary comb generation (MI) can occur for sufficiently high external pump intensities in both the normal and the anomalous dispersion regime, and shown examples of stable soft excitations frequency combs where the comb spectrum has been filled in. We have further shown examples of cascaded comb generation and stable background less anomalous dispersion frequency combs without a DC component that have been obtained for negative detuning. Numerical simulations indicate that combs with a DC component are generally unstable in the anomalous dispersion regime while stable for normal dispersion. We have additionally found that the dual pump configuration is capable of supporting cavity solitons. Potential benefits of using dual pumping for optical soliton memories or buffers is that interactions are suppressed, and that there is no background except for a low intensity pattern of clock pulses that can be controlled using the modulation frequency.
\vspace{.1cm}
\section*{Acknowledgements}

This research was funded by Fondazione Cariplo (grant no. 2011-0395), and the Italian Ministry of University and Research
(grant no. 2012BFNWZ2).

\end{document}